\documentclass[iop,twocolumn]{aastex63}

\usepackage{graphicx}
\usepackage{amssymb}
\usepackage{amsmath}
\usepackage{natbib}
\usepackage{longtable}
\usepackage{footnote}
\usepackage{helvet}
\usepackage{color}
\usepackage{afterpage}
\usepackage{float}
\usepackage{url}
\usepackage{enumitem}

\newcommand{\ayushiemail}{ayushi.singh@mail.utoronto.ca}

\newcommand{\cdmnew}{ } % { \color{magenta}}
\newcommand{\newtext}{ }   %{\color{magenta}}

\newcommand\T{\rule{0pt}{2.6ex}} % Top strut
\newcommand\B{\rule[-1.2ex]{0pt}{0pt}} % Bottom strut

\newcommand{\calTcl}{{\cal T}_{\rm cl}}
\newcommand{\calTbulk}{{\cal T}_{\rm bulk}}

\newcommand{\Sigmacl}{\Sigma_{\rm cl}}
\newcommand{\Sigmaenv}{\Sigma_{\rm env}}
\newcommand{\Sigmabg}{\Sigma_{\rm bg}}

\newcommand{\calWcl}{{\cal W}_{\rm cl}}
\newcommand{\calWtwod}{{\cal W}_{\rm cl,2D}}
\newcommand{\calTtwod}{{\cal T}_{\rm cl, 2D}}
\newcommand{\sigmaclz}{\sigma_{{\rm cl},z}}
\newcommand{\sigmacl}{\sigma_{{\rm cl}}}

\newcommand{\sigmaNT}{\sigma_{{\rm NT},z}}

\newcommand{\clumpnumber}{85 }

\newcommand{\Rcl}{R_{\rm cl}} 
\newcommand{\Mcl}{M_{\rm cl}} 
\newcommand{\vCMz}{v_{{\rm CM},z}} 
\newcommand{\alphaBM}{\alpha_{\rm BM92}}

\newcommand{\alphacl}{\alpha}% {\alpha_{\rm cl}}
\newcommand{\alphatwod}{\alpha_{\rm SMJ19}}

\begin{document}

%%%-------------- TITLE --------------%%%
\title{Are massive dense clumps truly sub-virial? A new analysis using Gould Belt ammonia data\footnote{Submitted to {\em ApJ} on June 1, 2021.}}

\correspondingauthor{Ayushi Singh}
\email{\ayushiemail}

\shorttitle{Are massive dense clumps truly sub-virial?}
\shortauthors{Singh et al.}

%%%-------------- AUTHORS --------------%%%
\author{Ayushi Singh}
\affiliation{David A.~Dunlap Department of Astronomy \& Astrophysics, University of Toronto, 50 St.~George Street, Toronto, ON, M5S 3H4 Canada}
\affiliation{Canadian Institute for Theoretical Astrophysics, University of Toronto, 60 St.~George St., Toronto, ON, M5S 3H8, Canada}

\author{Christopher D.~Matzner}
\affiliation{David A.~Dunlap Department of Astronomy \& Astrophysics, University of Toronto, 50 St.~George Street, Toronto, ON, M5S 3H4 Canada}

\author{Rachel K.~Friesen}
\affiliation{David A.~Dunlap Department of Astronomy \& Astrophysics, University of Toronto, 50 St.~George Street, Toronto, ON, M5S 3H4 Canada} 

\author{Peter G.~Martin}
\affiliation{Canadian Institute for Theoretical Astrophysics, University of Toronto, 60 St.~George St., Toronto, ON, M5S 3H8, Canada}

\author{Jaime E.~Pineda}
\affiliation{Max-Planck-Institut f{\"u}r extraterrestrische Physik, Giessenbachstrasse 1, 85748 Garching, Germany}

\author{Erik Rosolowsky}
\affiliation{Department of Physics, University of Alberta, Edmonton, AB, Canada} 

\author{Felipe Alves}
\affiliation{Max-Planck-Institut f{\"u}r extraterrestrische Physik, Giessenbachstrasse 1, 85748 Garching, Germany}

\author{Ana Chac{\'o}n-Tanarro}
\affiliation{Observatorio Astronómico Nacional (OAN-IGN), Alfonso XII 3, 28014, Madrid, Spain}

\author{Hope How-Huan Chen}
\affiliation{Harvard-Smithsonian Center for Astrophysics, 60 Garden St., Cambridge, MA 02138, USA}

\author{Michael Chun-Yuan Chen}
\affiliation{Queen's University, 99 University Ave, Kingston, ON,  K7L  3N6, Canada}

\author{Spandan Choudhury}
\affiliation{Max-Planck-Institut f{\"u}r extraterrestrische Physik, Giessenbachstrasse 1, 85748 Garching, Germany}

\author{James Di Francesco}
\affiliation{Department of Physics and Astronomy, University of Victoria, 3800 Finnerty Road, Victoria, BC,  V8P 5C2, Canada}
\affiliation{Herzberg Astronomy and Astrophysics, National Research Council of Canada, 5071 West Saanich Road, Victoria, BC, V9E 2E7, Canada}

\author{Jared Keown}
\affiliation{Department of Physics and Astronomy, University of Victoria, 3800 Finnerty Road, Victoria, BC,  V8P 5C2, Canada}

\author{Helen Kirk}
\affiliation{Herzberg Astronomy and Astrophysics, National Research Council of Canada, 5071 West Saanich Road, Victoria, BC, V9E 2E7, Canada}

\author{Anna Punanova}
\affiliation{Ural Federal University, 620002, Mira st.  19, Yekaterinburg, Russia }
%Max-Planck-Institut f\"ur extraterrestrische Physik, Giessenbachstrasse 1, 85748 Garching, Germany

\author{Youngmin Seo}
\affiliation{Jet Propulsion Laboratory, NASA, 4800 Oak Grove Dr, Pasadena, CA 91109, USA}

\author{Yancy Shirley}
\affiliation{Steward Observatory, 933 North Cherry Avenue, Tucson, AZ 85721, USA}

\author{Adam Ginsburg}
\affiliation{National Radio Astronomy Observatory, Socorro, NM 87801, USA}

\author{Stella S.~R.~Offner}
\affiliation{Department of Astronomy, University of Massachusetts, Amherst, MA 01003, USA}

\author{H{\'e}ctor G.~Arce}
\affiliation{Department of Astronomy, Yale University, P.O.~Box 208101, New Haven, CT 06520-8101, USA}

\author{Paola Caselli}
\affiliation{Max-Planck-Institut f{\"u}r extraterrestrische Physik, Giessenbachstrasse 1, 85748 Garching, Germany}

\author{Alyssa A.~Goodman}
\affiliation{Harvard-Smithsonian Center for Astrophysics, 60 Garden St., Cambridge, MA 02138, USA}

\author{Philip C.~Myers}
\affiliation{Harvard-Smithsonian Center for Astrophysics, 60 Garden St., Cambridge, MA 02138, USA}

\author{Elena Redaelli}
\affiliation{Max-Planck-Institut f{\"u}r extraterrestrische Physik, Giessenbachstrasse 1, 85748 Garching, Germany}
\affiliation{Dipartimento di Fisica \& Astronomia, Universita' degli Studi di Bologna, Viale Berti Pichat, 6/2, I - 40127 Bologna, Italy}

\collaboration{24}{(The GAS collaboration)} 

%\authorcomment1{submitted to {\em ApJ}, June 2021}
%\submitted{in {\em ApJ}, August 2021}

%%%-------------- ABSTRACT--------------%%%
\begin{abstract}
Dynamical studies of dense structures within molecular clouds often conclude that the most massive clumps contain too little kinetic energy for virial equilibrium, unless they are magnetized to an unexpected degree.  This raises questions about how such a state might arise, and how it might persist long enough to represent the population of massive clumps.   In an effort to re-examine the origins of this conclusion, we use  ammonia line data from the Green Bank Ammonia Survey and {\em Planck}-calibrated dust emission data from {\em Herschel} to estimate the masses and kinetic and gravitational energies for dense clumps in the Gould Belt clouds.  We show that several types of systematic error can enhance the appearance of low kinetic-to-gravitational energy ratios: insufficient removal of foreground and background material; ignoring the kinetic energy associated with velocity differences across a resolved cloud; and over-correcting for stratification when evaluating the gravitational energy.   Using an analysis designed to avoid these errors, we find that the most massive Gould Belt clumps harbor virial motions, rather than sub-virial ones.   As a byproduct, we present a catalog of masses, energies, and virial energy ratios for 85 Gould Belt clumps.   

\end{abstract}


%\tableofcontents
%%%-------------- INTRODUCTION --------------%%%
\section{Introduction} \label{S:introduction}

A key parameter in the study of molecular clouds and their substructures is the virial ratio: 
\begin{equation} \label{alpha_def} 
\alphacl \equiv \frac{2\calTcl}{ |\calWcl|}
\end{equation} 
Here $\calTcl$ is the object's kinetic energy in its center-of-mass frame, and $\calWcl$ is its self-gravitational energy. 
The value of $\alphacl$ as a diagnostic tool arises from the fact that it compares prominent opposing terms in the virial theorem \citep{ChandrasekharFermi53,McKee1992} -- that is, in the competition of forces (expressed as energies) that cause inward or outward accelerations of an object's radius (expressed as the trace of its moment-of-inertia tensor).  Because any significant imbalance leads to rapid change, there is good reason to expect that some chosen collection of dense interstellar structures is close to a state of equilibrium, at least in a statistical sense, {\newtext especially if the structures live for at least a single crossing time}.\footnote{ Ephemeral non-equilibrium states are also possible.  For instance, $\alphacl$ approaches 2 from below in asymptotic, non-rotating, unmagnetized free-fall \citep{BallesterosParedes18_OutOfVirial}, whereas $\alpha\gg2$ for explosive motions in excess of the escape velocity. }  {\newtext Insofar as this is true, } one can read the value of $\alphacl$ as an indication of the other, less easily observed, forces or energetic terms. A state in which  $\alphacl>1$ suggests the importance of the kinetic surface term that represents confinement by external thermal or turbulent pressure, or by the ram pressure due rapid inflow or outflow (see \citealt{Goldbaum11}), or due to colliding flows.  For example, $\alphacl= 1.683$ in the critical state of an isothermal, unmagnetized, pressure-bounded sphere \citep{Ebert55,Bonnor56}.   If $\alphacl\gg1$ then gravity is negligible in comparison to external pressure: {\newtext an equilibrium object in this state} is `pressure-confined'.

In contrast, $\alphacl <1$  indicates the importance of an additional positive term, corresponding to an outward force that opposes the combination of self-gravity and external pressure. Magnetic fields supply one such force, although for molecular clumps and cores, both Zeeman measurements \citep{Crutcher2012} and estimates based on the Davis-Chandraskhar-Fermi method \citep{MyersBasu21_MagFields} indicate median mass-to flux ratios about twice the critical value.  This implies that the quasi-static portion of the magnetic force is rarely sufficient to fully offset gravity.  The fluctuating portion of the magnetic energy is also limited in its impact, {\newtext as it tends to be in equipartition with the turbulent portion of $\calTcl$ \citep{McKee1992,Federrath16_Equipartition}.}  Another, often overlooked outward force is the momentum injected by protostellar outflows or photo-ionized regions when star formation is especially active. These must be included as an inner surface term when their kinetic energies are not included in $\calTcl$.   
The importance of this term is evident in the fact that the kinetic energy in protostellar outflows can be comparable to $\calTcl$ \citep[e.g.,][]{Graves10}.  However, their effect cannot overwhelm $2\calTcl$ in the virial theorem, for the simple reason that star-driven flows go on to stir turbulence within the medium \citep{Matzner02, NakamuraLi07, Matzner07}.  Similar arguments apply to stellar radiation forces \citep{McKee1992}, which in any case are only significant in the presence of vigorous massive star formation \citep{KrumholzMatzner09,MurrayQuataertThompson10,Raskutti16,JumperMatzner18}. 
Yet another often-overlooked term comes from the  gravity of matter outside the clump boundary {\newtext \citep{BallesterosParedes06_SixMyths}}; however this should be small compared to $\calWcl$ for clumps that are over-dense and not bounded by tidal forces. 

For these reasons it is difficult to envision a scenario in which  any collection of interstellar structures  would be strongly `sub-virial' -- that is, characterized by $\alphacl\ll 1$.

It is very puzzling, therefore, that observational studies of dense substructures within molecular clouds often find that $\alphacl$ is well below unity for the most massive of these objects (see for instance \citealt{Kauffmann2013}, \citealt{Urquhart14_ATLASGALclumps}, and  \citealt{Traficante18_Massive70micronClumpMotions, Traficante18_LarsonLaws}).  Although the selection of objects varies from one study to another (as does the specific correlation between estimates of $\alphacl$ and mass), the substructures in question are all molecular `clumps': objects intermediate in scale between molecular clouds and the compact `cores' from which individual star systems are born.  If massive molecular clumps are truly sub-virial, this has important implications for the initial conditions for star cluster formation.

Could the strongly sub-virial appearance of massive molecular clumps in fact  be an artefact of the way that observations are taken or interpreted?  \citet{TraficanteLeeHennebelle18_AlphaBias} advance one reason that it might be.  They point out that the data used to determine  $\calTcl$ and $\calWcl$ tend to weight different regions of a clump, due to the influence of a critical density on molecular line excitation, and that this may lead to a systematic offset in $\alphacl$. 

Here we explore another possibility: that choices involved in the method used to estimate $\alphacl$ may themselves introduce systematic errors. Because it depends on an object's three-dimensional density, temperature, and velocity fields, $\alphacl$ cannot be determined from projected data.  One must construct a proxy, such as the virial parameter introduced by \citeauthor{Bertoldi1992} (\citeyear{Bertoldi1992}, hereafter BM92) or the estimate presented by  \citeauthor{Singh2019} (\citeyear{Singh2019}, hereafter SMJ19).  We find that at several steps of this process, common analysis choices have the cumulative effect of suppressing the derived value of  $\alphacl$, especially for high-mass clumps.

For our exploration we employ 
NH$_3$ data from the Green Bank Ammonia Survey (GAS: \citealt{Friesen2017}) and column densities derived from a new analysis of dust optical depth in the {\em Herschel} data (A.~Singh \& P.G.~Martin, in prep.).  These provide sensitive, uniform, well-calibrated, and well-resolved information for an investigation such as ours. As a byproduct, we present a catalog of  properties for \clumpnumber clumps in the Gould Belt, 
using these  high-quality data.  Our catalog overlaps previous virial analyses, based on a subset of the same data, conducted by \citet{Kirk2017}, \citet{Redaellli17_B59}, \citet{Keown2017}, \citet{Chen19droplets}, and \citet{Kerr19virial}. However, we use a somewhat different algorithm to define clump boundaries; this allows us to focus on the  influence of analysis choices on estimates of $\alphacl$. 

In \S~\ref{S:method} we review  methods for estimating $\alphacl$. We introduce the data for our study in \S~\ref{S:data}, and present the details of our technique in \S~\ref{S:Implementation}. As a case study, we highlight the analysis of a single clump in \S~\ref{SS:CaseStudy}.   In \S~\ref{S:Results} we present results for our full sample of Gould Belt clumps.  Finally, in \S~\ref{S:conclusion}, we draw conclusions about the impact of biases on the apparent physical state of massive clumps. 

%%%-------------- METHOD --------------%%%
\section{Methods} \label{S:method}

The most widely used method for estimating $\alphacl$ involves the `virial parameter' introduced by BM92: 
\begin{equation} \label{eq:alphaBM_def}
\alphaBM = \frac{5 \sigmacl^2 \Rcl}{G\Mcl}. 
\end{equation}
Here $\Mcl$ and $\sigmacl$ are the mass and  one-dimensional velocity dispersion, respectively, of the object under study (a clump, in our case), and $\Rcl$ is its effective radius, usually defined so the projected area of the clump is $\pi \Rcl^2$.  We adopt that definition as well. 

The value of $\alphaBM$ derives from the fact that the gravitational energy of an interstellar object is usually similar to that of a uniform sphere with the same mass and effective radius; hence the correction factor $a$, defined by $\calWcl = -3 a G \Mcl^2/5\Rcl$, is of order unity.  BM92 consider the class of spheroidal clumps with power-law density profiles $\rho\propto r^{-k}$ in spheroidal radius coordinate $r$ as an example.  For these, $a$ can be decomposed ($a = a_1 a_2$) into a stratification factor $a_1 = (1-k/3)/(1-k/2.5)$, and a geometric factor $a_2$ whose angle average is close to unity except in the case of very prolate spheroids (see  Figure 2 of BM92). 

Along with the definition of $a$,  equations (\ref{alpha_def}) and (\ref{eq:alphaBM_def}) imply 

\begin{equation} 
\alphacl = \frac{\alphaBM}{a}\left(\frac{2\calTcl}{3\Mcl\sigmacl^2}\right),
\end{equation} 
showing that either $\alphaBM$, or the refined quantity $\alphaBM/a$, can be used to estimate $\alphacl$.  An important detail is that $\sigmacl$ must be defined so that  $3\Mcl \sigmacl^2$ is a valid estimate for $2\calTcl$; we return to this point below. 

As an alternative to $\alphaBM$ we will consider the method proposed by SMJ19, who   
%In a recent contribution,
re-examine the process of estimating  $\alphacl$ and suggest a procedure that is robust for non-spheroidal structures despite the effects of projection. 

The SMJ19 method begins with how a clump is identified and extracted from projected data.  One must start by defining a projected clump boundary, which encompasses a peak in the column density $\Sigma$.  The projected clump boundary should also contain reliable molecular line data from which to obtain the line-of-sight radial velocity $v_z$, the one-dimensional thermal velocity dispersion $\sigma_{\rm th}$,  and the total line-of-sight velocity dispersion $\sigma_z$, from which the non-thermal portion $\sigmaNT$ can be derived.  

The next step is to extract the cloud or clump column density $\Sigmacl$ from any external material projected within the cloud boundary.  Foreground and background removal is especially important when $\Sigma$ is derived from submillimeter dust emission; note that  spatially filtered observations (such as chopped or interferometric ones) accomplish this in an approximate way.  SMJ19 demonstrate that an approximation based on the Abel transform is more successful than either keeping all material within the cloud boundary, or using simple interpolation to clip a background level.  Abel reconstruction uses the fact that there is an exact relationship between any axisymmetric density distribution and its projection.  Applied to the column density, it provides an estimate for the component interior to a three-dimensional surface whose projection is the two-dimensional clump boundary.  The {\newtext removed component, $\Sigmaenv = \Sigma-\Sigmacl$,} is always lower toward the core of a clump than toward its edge, thanks to a projection effect that is analogous to limb brightening.  

As only one component of the cloud's kinetic energy is visible in projection, the quantity

\begin{equation} \label{eq:calT2D} 
\calTtwod =\calTbulk +  \frac32 \int \sigma_{\rm eff}(x,y)^2\Sigmacl(x,y)\,
dx \,dy,
\end{equation} 

where $x$ and $y$ are sky coordinates at the cloud distance,  provides an estimate for $\calTcl$ that is unbiased, in the sense that its average over viewing angles (denoted $\left<\cdots\right>$) is exact:  

\begin{equation} \label{eq:avgEkin}
\left<\calTtwod\right>=\calTcl.
\end{equation}  

We note that BM92 adopt the same quantity, defining $\sigmacl^2 = 2\calTtwod/3\Mcl$ in their Appendix C.   In equation (\ref{eq:calT2D}), 
\[ \sigma_{\rm eff}(x,y)^2 = \sigma_{\rm th}(x,y)^2 + \sigmaNT(x,y)^2\] is the square of the  effective one-dimensional velocity dispersion along each line of sight, and 

\begin{equation}
    \calTbulk = \frac32 \int [v_z(x,y)-\vCMz]^2\, \Sigmacl(x,y)\, dx\,dy
\end{equation}

is the contribution from resolved variations in the line-of-sight velocity $v_z$ across the cloud, which we will refer to as `bulk' kinetic energy.  Note that this may include a portion due to rotation. The center-of-mass velocity $\vCMz$ is computed from $\Mcl\vCMz =\int \Sigmacl(x,y) v_z(x,y) dx\,dy$.  

It is worth noting that $\calTbulk$ will be resolution-dependent, in practice, because $\sigmaclz(x,y)$ can only be defined at the resolution of a given experiment.  In the limit that a cloud is too small to be resolved, it would be described by the single centroid velocity  $v_z = \vCMz$, and thus have $\calTbulk=0$.  However, its observed kinetic energy should still be captured well by equation (\ref{eq:calT2D}), because its line width $\sigmaclz$ will include these unresolved velocity gradients. 

For the denominator of $\alphacl$, SMJ19 define a quantity $\calWtwod$, derived from $\Sigmacl$, with the desirable property that 

\begin{equation} \label{eq:avgEgrav}
\left<\calWtwod\right> = \calWcl.
\end{equation}

SMJ19 show that $\calWtwod$ can be computed by collapsing the clump mass profile to a sheet in the plane of the sky, obtaining the sheet's gravitational self-energy, and correcting the result by a factor $2/\pi$.  By using information from the resolved column density map, this procedure avoids the need to choose $\Rcl$ and  estimate $a$.  It therefore provides a means to calibrate the characteristic value of $a$ for an ensemble of clouds. 

With these definitions for $\calTtwod$ and $\calWtwod$, SMJ19's quantity 

\begin{equation} \label{eq:alpha2d_def}
 \alphatwod = \frac{2 \calTtwod}{|\calWtwod|}
\end{equation}   

is a valid estimate for $\alpha$, in the sense that both the numerator and denominator are exact when averaged over viewing angles. 

It is important to note that  $\alphatwod$ and $\alphaBM/a$ are equivalent if evaluated under the following conditions: (1) they are derived from the same model of the clump column density profile $\Sigmacl$ (and furthermore, to give valid estimates of $\alphacl$, this must have been been cleaned of foreground and background contamination); (2) the quantity $\sigmacl$ is defined so that $2 \calTtwod = 3 \Mcl \sigmacl^2$, thus incorporating $2\calTbulk$; and (3) the BM92 correction factor $a$ is evaluated in a way that properly reflects the clump profile. 

As we shall see, other choices for $\Mcl$, $\sigmacl$, and $a$ tend to introduce biases when  $\alphaBM/a$ (sometimes called $M_{\rm vir}/\Mcl$) is used as an estimate for $\alpha$.  

%%%-------------- DATA --------------%%%

\section{Data} \label{S:data}

\subsection{GAS molecular line data} \label{SS:gas}
All of the thermal and kinetic information we use in the  calculation of $\calTtwod$ derives from ammonia line data from the Green Bank Ammonia Survey (\citealt{Friesen2017}).  GAS observations of the NH$_3$ (1,1) and (2,2) inversion transitions, carried out with the K-Band Focal Plane Array of the Green Bank Telescope, achieve sufficient sensitivity ($\sim$ 0.1\,K median noise), spatial resolution (32", or 0.047 pc at 300 pc), and frequency resolution (5.7\,kHz/23.7\,GHz, or 0.07\,km\,s$^{-1}$) to map and resolve gas temperature, centroid velocity, and velocity dispersion across a wide range of dense core, clump, and filamentary structures. 
These  parameters  are  obtained  from  a  single-component  fit, a point we return to in \S \ref{SS:error}.

Our source data (GAS Data Release 2: J.\ Pineda et al., 2021, in prep.)\ incorporate the latest improvements to the GAS analysis pipeline relative to Data Release 1, including improved sensitivity arising from changes to the multi-component fitting pipeline, as well as an expanded data set of star-forming regions. The regions examined are listed in Table \ref{table:region_names}.

\begin{deluxetable}{lcc}[ht]
 \label{table:region_names}
\tabletypesize{\footnotesize}
\tablecolumns{5} 
\tablewidth{0pt}
 \tablecaption{ Cloud regions and adopted distances}
 \tablehead{
 \colhead{Cloud} & \colhead{Herschel Name}&\colhead{Adopted Distance}\\
 &&\colhead{[pc]}
 } 
\startdata 
B1 & Perseus & 301$^{\rm a}$%\tablenotemark{a}
\\
L1448 & Perseus & 288$^{\rm a}$%\tablenotemark{a}
\\
L1451 & Perseus & 279$^{\rm a}$%\tablenotemark{a}
\\
L1455 & Perseus & 235$^{\rm b}$\\
NGC1333 & Perseus & 293$^{\rm c}$%\tablenotemark{b}
\\
Perseus & Perseus & 235$^{\rm b}$ \\
IC348 & Perseus & 320$^{\rm c}$%\tablenotemark{b}
\\
B18 & Taurus & 126.6$^{\rm d}$%\tablenotemark{c}
\\
HC2  & Taurus & 138.6$^{\rm d}$%\tablenotemark{c}
\\
IC5146 	& IC5146	& 831$^{\rm e}$%\tablenotemark{d}
\\
Cepheus L1228 & Cepheus & 346$^{\rm f}$ \\
Cepheus L1251 & Cepheus & 346$^{\rm f}$%\tablenotemark{e}
\\
%B59 & Pipe & 163$^{\rm d}$%\tablenotemark{d}
%\\
CrA west & Corona Australis & 154$^{\rm e}$%\tablenotemark{d}
\\
CrA east & Corona Australis & 154$^{\rm e}$%\tablenotemark{d}
\\
L1688 & Ophiuchus & 138.4$^{\rm g}$%\tablenotemark{f}
\\
L1689 & Ophiuchus & 144.2$^{\rm g}$%\tablenotemark{f}
\\
Serpens MWC279 & Serpens & 437$^{\rm h}$ \\
OrionA & Orion A & 388$^{\rm i}$%\tablenotemark{g}
\\
OrionA S & Orion A & 428$^{\rm i}$%\tablenotemark{g}
\\
OrionB NGC2023-2024 & Orion B & 420$^{\rm i}$%\tablenotemark{g}
\\
OrionB NGC2068-2071 & Orion B & 388$^{\rm i}$%\tablenotemark{g}
\enddata
 \tablecomments{$^{\rm a}$\citet{zucker2018}; $^{\rm b}$ \citet{Hirota08}; 
 $^{\rm c}$\citet{ortizleon2018b}; $^{\rm d}$\citet{galli2018}; $^{\rm e}$\citet{dzib2018}; $^{\rm f}$\citet{yan2019}; $^{\rm g}$\citet{ortizleon2018a}; $^{\rm h}$\citet{OrtizLeon17}; $^{\rm i}$\citet{kounkel2017}}
\end{deluxetable}

\subsection{Herschel-derived column densities} \label{SS:column}

To estimate the  mass column density we employ  dust optical depth maps derived by fitting a spectral energy distribution (SED) to continuum data from the {\em Hershel} Space Observatory at 160, 250, 350 and 500 $\micron$. We use the results of an  improved analysis to be described in an upcoming work (A.~Singh \& P.G.~Martin, in prep.). 
A zero-point correction was applied to the Herschel intensity maps at each wavelength by correlating them with intensity models created from {\em Planck} dust models \citep{planck2013-p06b}.  Intensity maps were then fitted by a modified blackbody to estimate the dust temperature and optical depth using the dust emissivity index $\beta$ determined in each pixel from the {\em Planck} dust models. In the Singh \& Martin SED fitting pipeline, various cross-comparisons are used to optimize the determination of data and model uncertainties, thereby improving the robustness of the final maps.    We adopt a constant dust-plus-gas opacity of $\kappa_{\nu,0} = 0.1 $cm$^2$\,g$^{-1}$ at 1THz (\citealt{hildebrand83}, with a gas-to-dust mass ratio of 100) to determine $\Sigma$.

%%%-------------- ANALYSIS AND RESULTS --------------%%%
\section{Implementation} \label{S:Implementation}

%\subsection{Clump identification} \label{SS:region}

\begin{figure*}[t]
\centering
$\begin{array}{c}
	%\vspace{0.5cm}
    \hspace{-1.0cm}
    \includegraphics[scale=0.55]{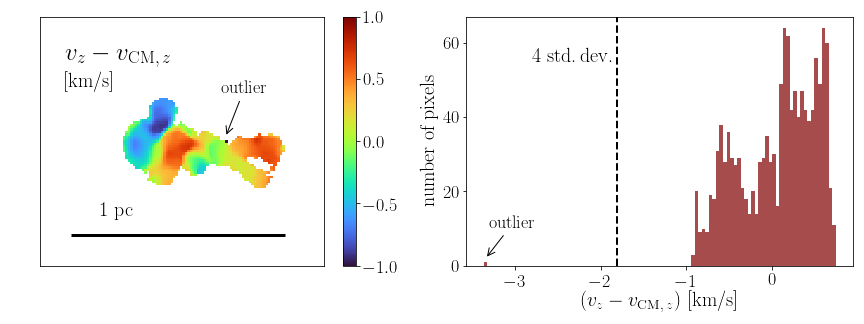}
\end{array}$
\caption{NH$_3$ line of sight velocity data for Clump 1 in Cepheus L1251, displaying a contaminating velocity outlier and our method for removing it.   The left panel shows $v_z$ relative to the center-of-mass velocity within the original clump boundary.  A single pixel of this map lies outside four standard deviations from the clump velocity distribution, as seen on the right panel.  We revise the cloud boundary to excise this pixel, and re-calculate quantities like $\vCMz$ before deriving $\calTcl$.  Variations of $v_z$ across the map contribute a  bulk kinetic energy $\calTbulk$ that can be important: in this case, $\alphatwod = 2.05$, but evaluates to 0.65 if $\calTbulk$ is ignored. }
\label{fig:bulk}
\end{figure*}

We identify objects for our analysis as identifiable peaks in NH$_3$ emission and total column density.  We draw projected clump and region boundaries around these according to the following procedure. 

To isolate structures for which we have complete and well-resolved data, we start by creating a  version of the NH$_3$ column density map that we convolve with a bounded parabolic (Epanechnikov) kernel of 88'' radius. 
Contours of this smoothed map are candidates for the clump boundary, from which we choose the largest for which two criteria are met: First, we must have complete data coverage to determine $\Sigma$, $v_z$, and $\sigma_z$, and nearly complete  coverage for $T_k$, within the boundary.  This requires that NH$_3$ (1,1) and (2,2) transitions are both well detected.  (We fill any gaps in the temperature data  using linear interpolation, which adds a negligible uncertainty to $\calTcl$.)  Second, we require that the inferred mass density $3\Mcl/4\pi\Rcl^3$ implies $n_H>3.6 \times 10^{3}$\,cm$^{-3}$, the critical density of the NH$_3$ (1,1) transition, which exceeds that of the (2,2) transition.  Because NH$_3$ has a reasonably consistent abundance at these densities \citep[e.g.,][]{Redaellli17_B59} and is not affected by freeze-out or significant line optical depth, this choice ensures that our kinematic information can be used to calculate $\calTcl$ in a reasonably unbiased way.  Nevertheless, abundance variations \citep[e.g.,][]{Crapsi07} do inevitably inject some uncertainty. 

Once a trial  boundary has been identified, it is modified if necessary to exclude regions in which $v_z$ falls more than four standard deviations outside its overall distribution for the clump.  Velocity outliers probably represent errors in the line fitting process, or possibly NH$_3$ emission from unrelated structures along the line of sight, and so it is important to exclude the spurious additional kinetic energy they would imply. We find that this approach affects only a small fraction of the clump area and mass (a few pixels at most).  Figure \ref{fig:bulk} provides one example, for clump L1251-1 in Cepheus. 

{\newtext As the figure shows, it is possible for our procedure to produce a clump boundary that includes a hole, indicating where NH$_3$ data is lacking.  L1251-1 has by far the largest such hole,\footnote{\newtext We exclude a clump in Barnard 59 from our sample, on the grounds that it lacks NH$_3$ data over an extensive region including the column density peak.} covering 2\% of its area. A number of other clumps have holes that span a few pixels; we indicate these cases with asterisks in Table \ref{table:ClumpAlphas}.  Because the lack of NH$_3$ data tends to coincide with a region of low to zero $\Sigmacl$, it makes no practical difference to $\calTtwod$ or $\calWtwod$ whether we fill these holes with interpolated data.  In computing radii we use the outer area of the mask; excluding the holes would only slightly lessen $\Rcl$.  } 

For every clump, we must also identify the boundary of an enclosing `region' for the purposes of removing dust emission from foreground and background material, thereby creating a map with compact support for Abel transform analysis.  While the exact choice of region boundary is not significant, it should be separated well enough from the clump boundary to contain the envelope physically associated to the clump, yet also close enough to the clump to sample a similar column density of background and foreground material (or at least, the component of this material on long spatial scales).  

\begin{figure*}[t]
\centering
$\begin{array}{c}
	%\vspace{0.5cm}
    \hspace{-0.5cm}
    \includegraphics[scale=0.5]{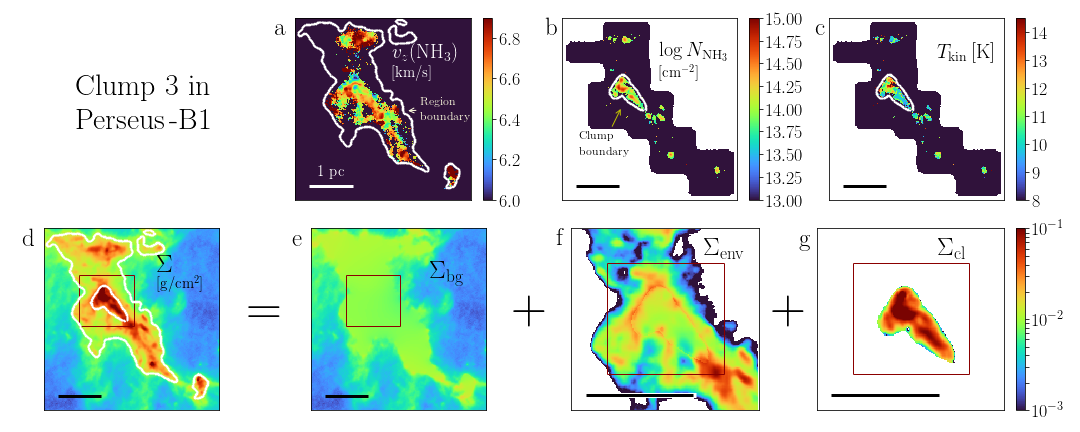}
\end{array}$
\caption{An example of the clump extraction process, applied to Clump 3 of the Perseus-B1 cloud discussed in \S\,\ref{SS:CaseStudy}.  Panels (a)-(c) show the molecular data for line-of-sight velocity, ammonia column density, and kinetic temperature, respectively, while (d)-(g), in turn, illustrate the total column $\Sigma$, removal of a foreground/background column $\Sigmabg$ (by interpolation), and the clump envelope contribution $\Sigmaenv$ (by Abel reconstruction), leaving $\Sigmacl$.  Note that the region boundary and clump boundary are identified in panels (a) and (b), respectively, as described in \S\,\ref{S:Implementation}.  Panels (f) and (g) are shown at higher magnification; the bar at the bottom of each panel indicates one parsec for scale. }

\label{fig:clumpExtration}
\end{figure*}

{To accomplish this, we generate a trial map and choose one of its contours to be the region boundary. We create a  version of the NH$_3$ column density smoothed with a Gaussian kernel of width $\Rcl/8$. We then multiply this by a vignetting function that is equal to the clump mask (unity within the clump boundary, zero outside), convolved with the function $1/(r_k+\Rcl)^{1/4}$ where $r_k$ is the distance from the centre of the kernel as measured in the plane of the sky. From this trial map, we choose the lowest contour that comes within $\Rcl/4$ of the clump boundary.  An example region boundary is displayed in {\newtext Figure \ref{fig:clumpExtration}.}  }  

Given the clump and region contours, we use the SMJ19 prescription to obtain $\Sigmacl$ in two steps.  First we subtract from $\Sigma$ a bicubic interpolation of its value from the region boundary; this is meant to extract an estimate of the emission from foreground/background dust.  We then apply the Abel transform, in the manner discussed by SMJ19, to subtract the emission from the clump's envelope, leaving $\Sigmacl$.  The lower panels of \ref{fig:clumpExtration} provides an illustration of the decomposition, for the case of clump 3 in Perseus-B1.  
   
From the clump column density profile $\Sigmacl$ we derive an estimate for  self-gravitational energy $\calWtwod$ according to  equation (4) of SMJ19; we ensure that $\calWtwod$ is accurate to within a fraction of one percent using numerical refinement and a correction for discretization errors, {\newtext which we describe in the Appendix.}   

In addition to our fiducial procedure, we consider several alternatives to identify what affects the trend of $\alphacl$ with $\Mcl$.  One such choice involves employing $\alphaBM/a$, but assuming a value for $a$ (such as unity).   Others involve evaluating $\Sigmacl$ differently.  For instance, one might replace the Abel transform with a simple interpolation, leaving a different clump column profile, or one might  adopt $\Sigmacl=\Sigma$, forgoing any foreground and background removal.  We shall also consider the effect of neglecting $\calTbulk$ when evaluating $\sigmacl$ and $\calTtwod$, to demonstrate that the velocity dispersion derived from individual beams does not capture the entire kinetic energy along the line of sight.

{\newtext Note that our definition of $\Rcl$, which derives from the projected area of NH$_3$ emission, differs from definitions involving fitted profiles or moments of the column density distribution. This should be kept in mind when comparing radii among catalogs. Moreover, because radius enters into the evaluation of $\alphaBM$, the method used to determine $\Rcl$ can introduce a new and potentially significant source of systematic error, which we do not attempt to evaluate here.   }

\subsection{Case studies: Cepheus L1251-1 and Perseus B1-3} \label{SS:CaseStudy}

As our primary case study we consider the clump shown in Figure \ref{fig:bulk}, Clump 1 within L1251 in Cepheus.    Applying our fiducial analysis we obtain $\calWtwod = -5.51\times 10^{44}$\,erg and $\calTtwod=5.75 \times 10^{44}$\,erg, of which $\calTbulk$ is 71\%.  These values imply $\alphatwod=2.05$.   

In Table \ref{table:case-study} we provide additional parameters for this particular clump, comparing  our fiducial (SMJ19) evaluation of $\alphacl$ against the BM92 analysis, and against alternative reconstructions of $\Sigmacl$ based on removing an interpolated column density level from the clump boundary (`Edge Interp.'), or making no correction for foreground and background contamination.  We also consider the effect of neglecting $\calTbulk$.

Several aspects of the analysis affect the result.  The method used to extract the clump's column density profile has a significant effect on its derived mass: simple interpolation from the clump boundary yields a 25\% lower estimate, whereas making no correction for foreground and background emission attributes 49\% more mass to the clump.  Of the observed kinetic energy, $\calTbulk$ constitutes  68\% in this case.  Equating $\calWtwod$ and $-3aG\Mcl^2/5\Rcl$ (or equivalently, $\alphatwod$ and $\alphaBM/a$) requires $a= 1.12$; notably, this is quite close to unity.  Taken together, we see that overestimating the mass, excluding $\calTbulk$, and adopting a value for $a$ significantly above unity (e.g., $a=a_1=5/3$, corresponding to $k=2$, as chosen by \citealt{RomanDuval10_RingGMCprops} and \citealt{Csengeri17_atlasgal}) can lead one to underestimate $\alphacl$ for this object by as much as a factor of 7.3.  

How representative is Cepheus L1251-1? We note that  Perseus B1-3, the clump depicted in  Figure \ref{fig:clumpExtration}, yields an identical estimate for $a$ (i.e., $a=1.12$) but contains  only 22\% of its kinetic energy in bulk form.  Our estimate for its virial ratio is $\alphatwod = 0.58$, the lowest of our entire sample.  In its case, neglecting to correct for foreground or background material, omitting $\calTbulk$, and adopting $a=5/3$ would lead one to underestimate $\alphacl$ by the more moderate factor of 2.7.  

In the next section we place these examples into context, and show that Cepheus L1251-1 and Perseus B1-3 are not outliers, but do bracket the range of massive clump properties.

\begin{deluxetable}{lccc}[h]
 \label{table:case-study}
\tabletypesize{\footnotesize}
\tablecolumns{5} 
\tablewidth{0pt}
 \tablecaption{Impact of analysis choices for  Cepheus L1251-1}
 \tablehead{
 \colhead{Quantity} & \colhead{Abel }&\colhead{Edge Interp.}&\colhead{No bg./fg.\ corr.}
 } 
\startdata 
$\Mcl$ $[M_\odot]$ & 56.2& 43.2 & 82.6 \\
$\Rcl$ [pc] & 0.32& 0.32 & 0.32 \\
$\sigmacl$ [km/s] & 0.59& 0.6 & 0.58 \\
$|\calWtwod|$ [erg] & $ 5.68 \times 10^{44}$ & $ 3.62  \times 10^{44}$ & $ 1.08 \times 10^{45}$\\
$\calTtwod$  [erg]& $ 5.83  \times 10^{44}$ & $ 4.59 \times 10^{44}$& $ 8.2  \times 10^{44}$\\
$\calTbulk$  [erg]& $ 3.99  \times 10^{44}$ & $ 3.16 \times 10^{44}$& $ 5.55  \times 10^{44}$\\
$\alphatwod$ & 2.05& 2.54 & 1.52\\
$\alphatwod$ (No $\calTbulk$) & 0.65& 0.79 & 0.49\\
$\alphaBM$ & 2.3& 3.05 & 1.49 \\
$\alphaBM$ (No $\calTbulk$) & 0.73& 0.95 & 0.48
\enddata
\tablecomments{Columns indicate method used to remove foreground and background emission from $\Sigmacl$: our fiducial Abel-transform reconstruction, interpolation from the clump boundary, or no correction at all.  Rows marked `No $\calTbulk$' indicate that resolved kinetic energy is omitted from $\calTcl$ and $\sigmacl$. }
\end{deluxetable}

\section{Results: are massive dense clumps truly sub-virial?} %\label{SS:virialratio}
\label{S:Results}

We list our findings for all \clumpnumber of our Gould Belt NH$_3$ clumps in Table \ref{table:ClumpAlphas}, and plot them for various choices of the analysis method in Figures \ref{fig:smj_bm} and \ref{fig:abel_edge}.  

In Figure \ref{fig:smj_bm} we adopt our fiducial method (Abel reconstruction) for removing foreground and background matter, and compare the outcomes of other choices in the estimation of $\alphacl$.
Two trends are immediately clear.  First, $\alphaBM$ and $\alphatwod$ are very similar across our entire sample.  Because we use the same data to derive both quantities, the comparison provides a  statistical calibration of BM92's self-gravity parameter $a$, which our analysis shows is only slightly above unity (geometric mean value 1.125) in our Gould Belt clumps.  For more detail, examining the ratio of the fits plotted in Figure \ref{fig:smj_bm}, we see that $a$ grows from unity to a maximum of 1.15 across the range of $\Mcl$. 

Second, excluding $\calTbulk$ suppresses $\alphacl$ by an amount that changes with $\Mcl$.  The difference is insignificant for $\Mcl<M_\odot$, where clumps harbour subsonic motions and are not highly resolved.  However, the {\cdmnew influence of $\calTbulk$} grows with mass and amounts to  0.26\,dex (a factor of 1.81) for $\Mcl \sim 100\,M_\odot$.   This is important, as most analyses adopt for $\alphacl$ a typical value of the total beam-wise line width $\sigma_{\rm eff}(x,y)$, which has the effect of neglecting $\calTbulk$ in the estimate of $\calTtwod$.  (There do exist counter-examples, however: for instance, \citealt{RomanDuval10_RingGMCprops}'s equation 7 includes all of $\calTtwod$.)   

{\cdmnew Methods for extracting $\Sigmacl$ from the column density map $\Sigma$ also influence the virial parameter. In Figure  \ref{fig:abel_edge},  we compare several methods: removing emission from around the clump using our default method of Abel reconstruction, removing an interpolated value from the clump boundary,  or making no correction (in which case $\Sigmacl=\Sigma$ within the boundary). } The trends in $\alphacl(\Mcl)$ are shown for these different choices.

Simple interpolation removes the most mass, while Abel reconstruction, by accounting for projection effects, attributes less material to the envelope.  These differences are greatest for the least massive clumps, whose column density contrasts the least with their surroundings.   The choice of extraction method  affects the inferred virial parameter, mostly through the relation $\alphacl\propto \Mcl^{-1}$ that applies when $\Sigmacl$ is multiplied by a constant. However,  the effect on $\alphacl$ diminishes for higher clump masses.  The  effect is also roughly in the direction of the trend line $\alphacl(\Mcl)$ --  that is, it is partly degenerate with the original trend.  These facts diminish the influence of the extraction  method on the trend of  $\alphacl$ with clump mass.  

In summary: although $\alphacl$ drops with $\Mcl$ within our collection of clumps, it is consistently below unity at high masses only  when one ignores spatially resolved component of the bulk kinetic energy,  when one overestimates the gravitational self-energy by adopting $a>1.1$, {\cdmnew or when one does not correct at all for foreground and background matter projected within the clump boundary}.  Using our most complete analysis, which addresses these shortcomings, we find no evidence that {\cdmnew the massive clumps in our sample} are consistently sub-virial. 

Note that the clumps in our sample typically have $\alphatwod\sim 5$ for $\Mcl=1\,M_\odot$, while studies of dense cores, such as \citet{Johnstone00_BEclumps}, find that cores approach the critical state of a Bonnor-Ebert sphere (in which $\alphacl=1.683$) at around a solar mass. 
The discrepancy is not surprising, considering that our algorithm defines clumps that include as much NH$_3$ emission as possible.  Any compact cores in our sample are therefore enclosed within larger clumps.   

\subsection{Sources of error} \label{SS:error}

\begin{figure*}[ht]
\centering
$\begin{array}{c}
	%\vspace{0.5cm}
    \hspace{0.0cm}
    \includegraphics[scale=0.65]{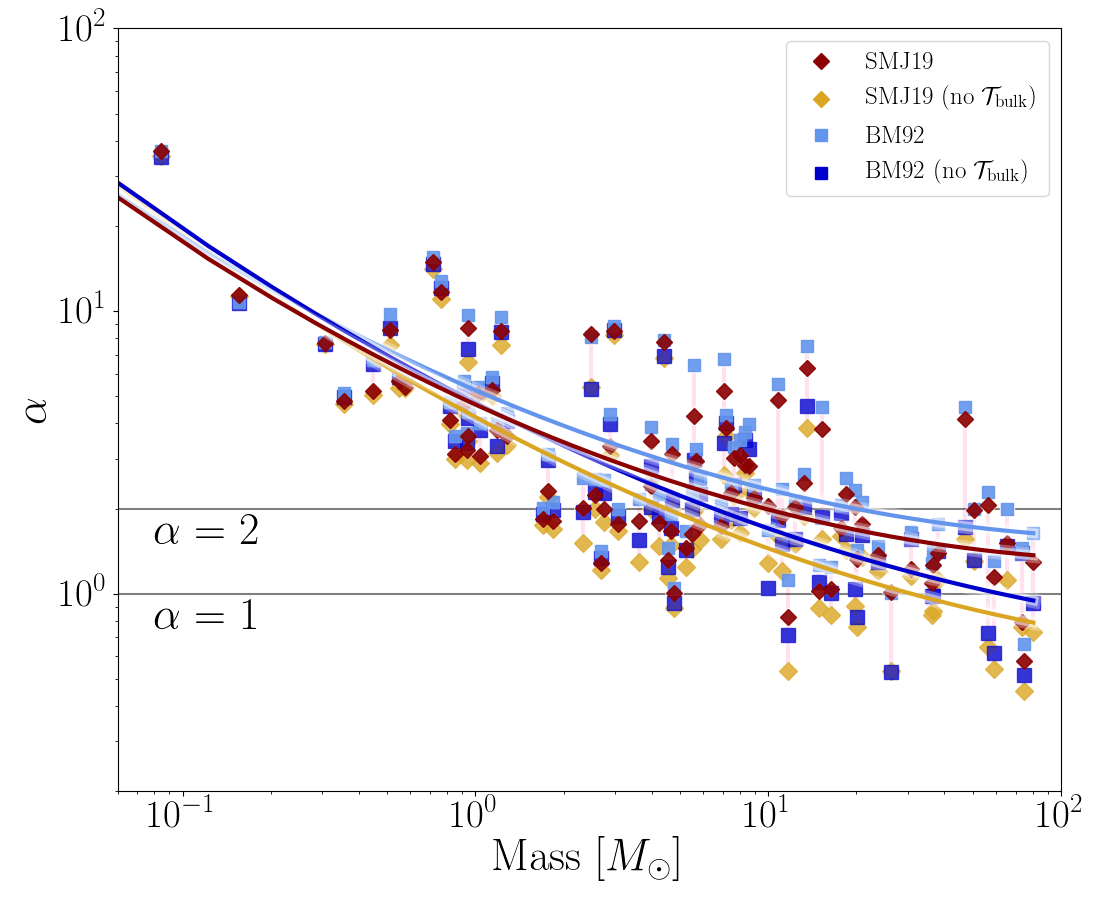}
\end{array}$
\caption{Dark and light blue squares $\alphaBM$ with and without the contribution of $\calTbulk$, respectively; red and orange diamonds represent $\alphatwod$ with and without $\calTbulk$, respectively.   For $\alphaBM$ we set $\Rcl$ equal to the geometric mean of semi-major and semi-minor axes of an ellipse fit to the clump boundary.  Pink lines connect data for the same clump.   The over-plotted curves are quadratic fits to $\log\alpha\,(\,\log\Mcl\,)$ for the points that share their line color, and are shown only to clarify trends among the four point families. }
\label{fig:smj_bm}
\end{figure*}

\begin{figure*}[ht]
\centering
$\begin{array}{c}
	%\vspace{0.5cm}
    \hspace{0.0cm}
    \includegraphics[scale=0.65]{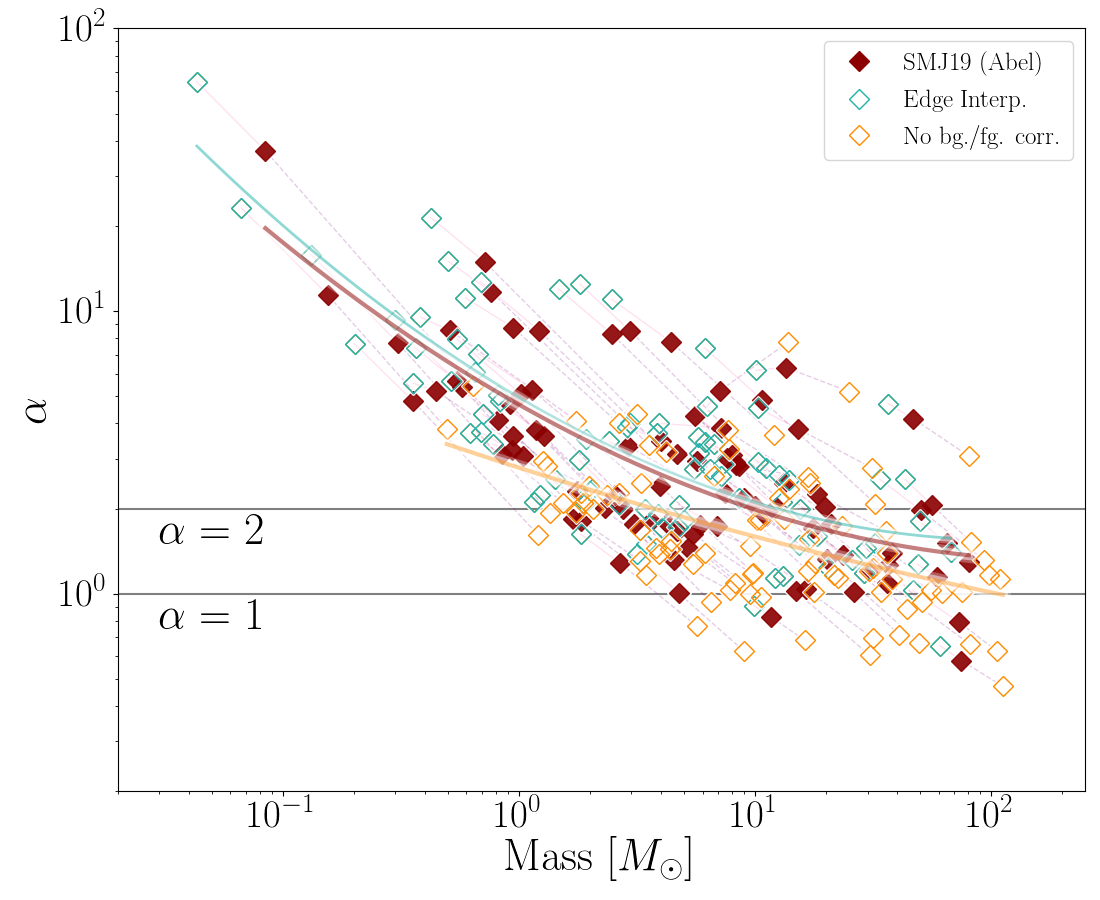}
\end{array}$
\caption{A demonstration that the method used to extract the clump column $\Sigmacl$ from the column density map $\Sigma$ affects the  derived virial ratio-mass relation.  Solid red diamonds represent our fiducial Abel reconstruction, as in Figure \ref{fig:smj_bm}; open cyan diamonds represent the removal of a bi-cubic interpolation from the cloud edge; and open orange symbols indicate no correction (all dust emission projected within the clump boundary is attributed to $\Sigmacl$).  Pink solid and purple dashed lines connect values for the same clump.  The analysis otherwise follows SMJ19. Over-plotted curves are quadratic fits in log space to the point families of the same color.  }  
\label{fig:abel_edge}
\end{figure*}

We pause here to discuss several sources of error,  approximately in descending order of their expected importance for our conclusions. 
\begin{itemize}[label=-]
  \setlength\itemsep{0.8em}
  
\item Despite being well-calibrated, our estimated column densities are subject to systematic errors arising from the assumed dust emissivity and dust-to-gas ratio.
Our assumptions that $\beta$ is constant along each line of sight, and of a universal ratio between dust optical depth and column density,  both introduce systematic uncertainties into our determination of $\Sigmacl$.  Variations of the dust opacity have primarily been observed in  densest regions of compact cores and filaments \citep[e.g.,][]{ChaconTanarro19_DustOpacVarL1544,Howard19_L1495_PPMAP}, implying that the impact of line-of-sight variations on our clump column densities is minor but not negligible.   {\newtext We note that the range of inferred $\beta$ values within each of our clumps is extremely limited, in that the median clump only spans a range $\Delta \beta = 0.005$ and the largest variation within a clump is $\Delta \beta = 0.051$. A caveat, considering the limited resolution of {\em Planck}, is that these values could be  significantly underestimated.  Overall, however,} we consider submillimeter dust emission a more reliable tracer than NH$_3$ for determining column density,  so long as the clump profile can be separated from other emission along the line of sight.

\item Our analysis, which is optimized for well-resolved clumps,  does not allow an explicit correction for finite resolution of the type introduced by \citet{RosolowskyLeroy06}. In this regard, it is useful to note that the clump angular diameter $2\Rcl$/distance correlates with clump mass, ranging from 44" to 8' across our sample, with median values of 2.8' and 3.8' for those clumps with $\Mcl<10\,M_\odot$ and $\Mcl>10\,M_\odot$, respectively. Comparing to the  GAS resolution of 32", we infer that finite resolution is likely to affect clump selection and to bias our estimates of $\alphacl$, but predominantly for the low-mass clumps that are not our primary focus. 

\item  We work only with projected data and line-of-sight velocities.  As was discussed in SMJ19, this leads to errors in both the numerator and denominator of $\alphatwod$ that are random insofar as the viewing angle is random (but become systematic if clumps are aligned in a larger structure, as \citealt{Alves20} find). These errors are greater for anisotropic structures and velocity fields, and so they tend to be greater for more massive clumps. Projection-dependent errors are surely responsible for some of the scatter we see in Figure \ref{fig:smj_bm}, which amounts to 0.21 dex standard deviation around the fitted curves. Notably, this is comparable to variation with viewing angle in the simulation examined by SMJ19.   One might be able to calibrate this component of the scatter using a statistical sample of simulated clumps, although that would be beyond the scope of the current work. 

\item We estimate $\calTtwod$ using 
parameters obtained from a single-component fit to the NH$_3$ emission, which could miss some kinetic energy from other components that represent fluid velocities within the clump.  Althogh  some  differences  are indeed seen  when single-comonent and multiple-component fits are compared  \citep{Choudhury20_UbiqNH3inL1688,Choudhury21_L1688_Transition},  these amount  to  small  corrections  for the clump kinematics.  Moreover, at GAS sensitivity, most pixels are well fit with single components \citep[e.g.,][]{Sokolov20_ProbabilisticLines}.

\item To remove foreground and background dust emission from $\Sigmacl$ requires deprojection, which necessarily implies some error \citep{Beaumont2013}.   SMJ19 found that our fiducial method based on the Abel transform performs better than either using simple interpolation or making no correction.  However, it probably implies both random and systematic errors.    Because the choice of method has a greater impact at lower clump masses, we infer that these errors declines with $\Mcl$ and are quite minor for the massive clumps of greatest interest here.

\item Although we require that our clumps exceed the critical densities of the NH$_3$ (1,1) and (2,2) transitions, excitation variations like those seen by \citet{Crapsi07} are likely to add biases to the data we use to estimate $\calTtwod$ and hence $\alphacl$, as argued by  \citet{TraficanteLeeHennebelle18_AlphaBias}. 
We also note that NH$_3$ data could be contaminated by emission projected within the clump boundary, which would cause $\alphacl$ to be slightly overestimated \citep[e.g.,][]{Choudhury20_UbiqNH3inL1688}.

\item We omit any component not traced by dust and NH$_3$ emission, such as embedded stars and high-velocity outflows.  In the case of protostellar outflows, this means that our definition of $\calTcl$ applies only to matter at velocities within a few $\sigmaclz$ of the systemic velocity, so that outflows must be treated as surface terms within the virial theorem (as we discussed in \S \ref{S:introduction}).  An alternative definition of $\calTcl$ would explicitly include the outflow kinetic energy.   In the case of protostars, our definition means that $\calWcl$ lacks the contribution from  stellar gravity. {\newtext  As the two effects both correlate with star formation, we expect them to be most important for the most massive, lowest-$\alphacl$ clumps in our sample.  If included in $\alphacl$, we expect they would affect it in opposite ways, with the positive effect of outflows being more significant than the negative effect of stellar gravity.  In the active region NGC~1333, for example, $\sim 20\%$ of the total mass is in protostars \citep[][see \citealt{Matzner2015}]{Gutermuth08_N1333protostars}, a value typical of embedded protoclusters \citep{LadaLada03_EmbeddedClusters}, while the protostellar outflow energy is comparable to $\calTcl$ \citep{Plunkett13_N1333outflows}. }  

\item Our method for evaluating gravitational energy converges when $\Sigmacl$ is well resolved {\newtext (as discussed in the Appendix)}, but underestimates the magnitude of  $\calWtwod$  when there exists unresolved structure.   However, the fact that $a$ is very close to its value for a uniform sphere is strong indication that the gravitational energy in unresolved structure is very minor. 
\end{itemize}

Of these, the systematic errors will affect our conclusions regarding $\alphacl(\Mcl)$, while random errors will tend to average out. 

It is important to note, however, that the conclusions we draw when comparing analysis choices are independent of measurement error, because we use identical data to make these comparisons. 

\startlongtable %%% <-- this allows the table to break across pages! 
\begin{deluxetable*}{lcccccccc}
\tabletypesize{\footnotesize}
\tablecolumns{9} 
\tablewidth{0pt}
 \tablecaption{ Clump properties and virial ratios derived from Abel reconstruction }
 \tablehead{
 \colhead{Clump} & \colhead{RA} & \colhead{DEC} &  \colhead{Mass} &\colhead{Radius}&\colhead{$\alphatwod$ } & \colhead{$\alphatwod$ } & \colhead{$\alphaBM$} & \colhead{$\alphaBM$} \\
& &&\colhead{[M$_{\odot}$]} &[pc] && (no $\calTbulk$) && (no $\calTbulk$)
 } 
\startdata 
B1 1 	& 03h33m36s 	& +31d18m49s 	& 2.97 	& 0.11 	& 8.49 	& 8.23 	& 8.82 	& 8.55 \\
B1 2 	& 03h33m18s 	& +31d18m31s 	& 2.89 	& 0.13 	& 3.34 	& 3.08 	& 4.32 	& 3.99 \\
B1 3 	& 03h33m18s 	& +31d04m44s 	& 74.93 	& 0.31 	& 0.58 	& 0.45 	& 0.66 	& 0.51 \\
B1 4 	& 03h32m43s 	& +30d58m41s 	& 12.32 	& 0.21 	& 2.03 	& 1.50 	& 2.12 	& 1.57 \\
B1 5 	& 03h32m24s 	& +30d48m01s 	& 5.48 	& 0.11 	& 1.62 	& 1.44 	& 2.25 	& 2.00 \\
B1 6 	& 03h31m26s 	& +30d44m00s 	& 3.99 	& 0.10 	& 3.47 	& 2.52 	& 3.89 	& 2.83 \\
L1448 1 	& 03h25m41s 	& +30d42m49s 	& 58.84 	& 0.28 	& 1.15 	& 0.54 	& 1.30 	& 0.61 \\
L1451 1 	& 03h25m35s 	& +30d20m17s 	& 2.56 	& 0.11 	& 2.23 	& 1.99 	& 2.55 	& 2.28 \\
L1451 2* 	& 03h24m33s 	& +30d22m09s 	& 0.16 	& 0.03 	& 11.35 	& 11.32 	& 10.74 	& 10.71 \\
L1455 1 	& 03h27m50s 	& +30d10m00s 	& 18.41 	& 0.25 	& 2.25 	& 1.42 	& 2.56 	& 1.63 \\
L1455 2 	& 03h28m07s 	& +30d04m55s 	& 3.97 	& 0.14 	& 2.40 	& 1.99 	& 2.44 	& 2.02 \\
L1455 3 	& 03h27m36s 	& +29d57m10s 	& 0.55 	& 0.06 	& 5.64 	& 5.34 	& 6.17 	& 5.85 \\
NGC1333 1 	& 03h30m00s 	& +31d37m39s 	& 2.76 	& 0.12 	& 1.99 	& 1.79 	& 2.52 	& 2.26 \\
NGC1333 2 	& 03h29m29s 	& +31d34m32s 	& 0.82 	& 0.07 	& 4.12 	& 3.97 	& 4.78 	& 4.60 \\
NGC1333 3 	& 03h29m29s 	& +31d31m51s 	& 1.77 	& 0.09 	& 2.31 	& 2.20 	& 3.11 	& 2.97 \\
NGC1333 4 	& 03h29m28s 	& +31d26m37s 	& 7.44 	& 0.20 	& 2.27 	& 1.82 	& 2.37 	& 1.91 \\
NGC1333 5 	& 03h29m05s 	& +31d18m36s 	& 20.97 	& 0.19 	& 1.76 	& 1.35 	& 2.11 	& 1.61 \\
NGC1333 6 	& 03h29m10s 	& +31d12m55s 	& 80.46 	& 0.29 	& 1.29 	& 0.73 	& 1.63 	& 0.93 \\
NGC1333 7 	& 03h28m42s 	& +31d13m29s 	& 15.24 	& 0.24 	& 3.83 	& 1.56 	& 4.58 	& 1.87 \\
NGC1333 8 	& 03h28m44s 	& +31d04m06s 	& 8.00 	& 0.17 	& 3.08 	& 1.64 	& 3.50 	& 1.86 \\
Perseus 1 	& 03h30m40s 	& +30d25m04s 	& 0.95 	& 0.06 	& 3.62 	& 3.46 	& 4.38 	& 4.18 \\
Perseus 2 	& 03h30m22s 	& +30d21m52s 	& 1.28 	& 0.10 	& 3.60 	& 3.37 	& 4.36 	& 4.08 \\
IC348 1 	& 03h45m23s 	& +32d03m24s 	& 0.92 	& 0.09 	& 4.68 	& 4.40 	& 5.67 	& 5.33 \\
IC348 2 	& 03h45m07s 	& +31d59m09s 	& 1.02 	& 0.09 	& 5.08 	& 4.95 	& 5.37 	& 5.23 \\
IC348 3 	& 03h44m57s 	& +31d59m03s 	& 1.14 	& 0.08 	& 5.26 	& 5.01 	& 5.84 	& 5.57 \\
IC348 4 	& 03h44m43s 	& +31d56m54s 	& 0.95 	& 0.07 	& 8.74 	& 6.63 	& 9.67 	& 7.34 \\
IC348 5 	& 03h44m26s 	& +31d57m31s 	& 5.68 	& 0.17 	& 2.96 	& 2.33 	& 3.25 	& 2.56 \\
IC348 6 	& 03h44m03s 	& +32d00m42s 	& 36.46 	& 0.28 	& 1.26 	& 0.87 	& 1.42 	& 0.98 \\
B18 1 	& 04h35m46s 	& +24d07m49s 	& 3.62 	& 0.08 	& 1.81 	& 1.29 	& 2.17 	& 1.55 \\
B18 2 	& 04h32m53s 	& +24d22m50s 	& 4.53 	& 0.09 	& 1.32 	& 1.13 	& 1.45 	& 1.25 \\
B18 3 	& 04h32m04s 	& +24d30m20s 	& 2.33 	& 0.08 	& 2.01 	& 1.52 	& 2.57 	& 1.94 \\
B18 4 	& 04h30m12s 	& +24d24m10s 	& 0.36 	& 0.04 	& 4.82 	& 4.69 	& 5.12 	& 4.98 \\
B18 5 	& 04h29m30s 	& +24d33m22s 	& 2.68 	& 0.06 	& 1.28 	& 1.21 	& 1.41 	& 1.33 \\
B18 6 	& 04h27m06s 	& +24d39m52s 	& 0.31 	& 0.05 	& 7.73 	& 7.62 	& 7.76 	& 7.64 \\
HC2 1 	& 04h41m39s 	& +25d59m46s 	& 4.77 	& 0.10 	& 1.00 	& 0.89 	& 1.05 	& 0.93 \\
HC2 2 	& 04h41m35s 	& +25d44m15s 	& 10.01 	& 0.16 	& 2.04 	& 1.28 	& 1.67 	& 1.05 \\
HC2 3 	& 04h40m38s 	& +25d28m03s 	& 0.08 	& 0.04 	& 36.92 	& 35.29 	& 36.74 	& 35.12 \\
HC2 4 	& 04h39m36s 	& +26d24m56s 	& 1.70 	& 0.07 	& 1.83 	& 1.75 	& 2.01 	& 1.92 \\
HC2 5 	& 04h39m24s 	& +25d50m26s 	& 0.51 	& 0.05 	& 8.53 	& 7.61 	& 9.76 	& 8.71 \\
HC2 6* 	& 04h39m45s 	& +25d39m39s 	& 1.18 	& 0.08 	& 3.81 	& 3.15 	& 4.03 	& 3.34 \\
IC5146 1 	& 21h47m31s 	& +47d31m17s 	& 65.38 	& 0.43 	& 1.51 	& 1.11 	& 2.00 	& 1.47 \\
IC5146 2 	& 21h47m15s 	& +47d31m37s 	& 7.19 	& 0.17 	& 3.87 	& 3.61 	& 4.28 	& 4.00 \\
IC5146 3 	& 21h46m08s 	& +47d34m15s 	& 17.70 	& 0.27 	& 1.71 	& 1.60 	& 2.06 	& 1.93 \\
IC5146 4 	& 21h45m14s 	& +47d32m02s 	& 23.69 	& 0.34 	& 1.37 	& 1.20 	& 1.48 	& 1.30 \\
IC5146 5 	& 21h45m06s 	& +47d38m21s 	& 38.07 	& 0.36 	& 1.39 	& 1.12 	& 1.76 	& 1.42 \\
Cepheus L1228 1 	& 20h58m07s 	& +77d42m03s 	& 3.07 	& 0.10 	& 1.76 	& 1.66 	& 2.00 	& 1.88 \\
Cepheus L1228 2 	& 20h57m09s 	& +77d39m51s 	& 0.57 	& 0.06 	& 5.39 	& 5.36 	& 5.90 	& 5.86 \\
Cepheus L1228 3 	& 20h57m44s 	& +77d34m26s 	& 11.11 	& 0.16 	& 1.88 	& 1.20 	& 2.34 	& 1.50 \\
Cepheus L1251 1* 	& 22h39m09s 	& +75d09m56s 	& 56.32 	& 0.32 	& 2.05 	& 0.65 	& 2.30 	& 0.73 \\
Cepheus L1251 2* 	& 22h36m15s 	& +75d17m16s 	& 4.25 	& 0.14 	& 1.77 	& 1.48 	& 2.26 	& 1.88 \\
Cepheus L1251 3 	& 22h30m55s 	& +75d12m09s 	& 26.25 	& 0.26 	& 1.01 	& 0.53 	& 1.00 	& 0.53 \\
Cepheus L1251 4 	& 22h28m25s 	& +75d12m29s 	& 4.66 	& 0.16 	& 1.67 	& 1.49 	& 1.91 	& 1.70 \\
CrAwest 1 	& 19h01m55s 	& -36d58m00s 	& 7.64 	& 0.09 	& 3.01 	& 2.22 	& 3.30 	& 2.43 \\
CrAeast 1 	& 19h10m26s 	& -37d09m45s 	& 1.84 	& 0.05 	& 1.81 	& 1.69 	& 2.11 	& 1.97 \\
L1688 1 	& 16h29m05s 	& -24d22m08s 	& 0.85 	& 0.05 	& 3.11 	& 3.00 	& 3.61 	& 3.48 \\
L1688 2 	& 16h28m38s 	& -24d19m58s 	& 0.94 	& 0.06 	& 3.23 	& 2.98 	& 3.70 	& 3.42 \\
L1688 3 	& 16h28m29s 	& -24d37m48s 	& 0.45 	& 0.05 	& 5.21 	& 5.03 	& 6.71 	& 6.48 \\
L1688 4 	& 16h28m06s 	& -24d35m03s 	& 1.04 	& 0.06 	& 3.08 	& 2.91 	& 4.02 	& 3.79 \\
L1688 5 	& 16h27m42s 	& -24d43m57s 	& 2.48 	& 0.09 	& 8.29 	& 5.39 	& 8.11 	& 5.28 \\
L1688 6 	& 16h27m26s 	& -24d29m39s 	& 13.28 	& 0.12 	& 2.47 	& 1.89 	& 2.64 	& 2.02 \\
L1688 7 	& 16h27m02s 	& -24d34m14s 	& 6.89 	& 0.12 	& 1.74 	& 1.56 	& 2.07 	& 1.85 \\
L1688 8 	& 16h26m35s 	& -24d25m00s 	& 5.83 	& 0.07 	& 1.75 	& 1.54 	& 2.54 	& 2.25 \\
L1689 1 	& 16h32m34s 	& -24d30m10s 	& 11.67 	& 0.09 	& 0.83 	& 0.53 	& 1.12 	& 0.72 \\
L1689 2 	& 16h31m48s 	& -24d51m46s 	& 5.23 	& 0.08 	& 1.45 	& 1.24 	& 1.66 	& 1.43 \\
L1689 3* 	& 16h32m03s 	& -24d58m59s 	& 4.68 	& 0.11 	& 3.12 	& 1.99 	& 3.38 	& 2.15 \\
Serpens MWC297 1 	& 18h28m15s 	& -03d48m49s 	& 16.44 	& 0.19 	& 1.03 	& 0.84 	& 1.24 	& 1.00 \\
OrionA 1 	& 05h35m11s 	& -04d57m12s 	& 8.95 	& 0.16 	& 2.19 	& 2.00 	& 2.42 	& 2.22 \\
OrionA 2 	& 05h35m11s 	& -05d37m50s 	& 46.92 	& 0.28 	& 4.15 	& 1.56 	& 4.59 	& 1.72 \\
OrionA 3 	& 05h35m14s 	& -05d53m21s 	& 7.09 	& 0.16 	& 5.22 	& 2.63 	& 6.76 	& 3.41 \\
OrionA 4 	& 05h36m17s 	& -06d12m15s 	& 8.60 	& 0.14 	& 2.82 	& 2.31 	& 3.97 	& 3.25 \\
OrionA 5 	& 05h35m17s 	& -06d15m18s 	& 13.53 	& 0.15 	& 6.31 	& 3.86 	& 7.54 	& 4.62 \\
OrionA S 1 	& 05h39m12s 	& -07d13m22s 	& 10.76 	& 0.22 	& 4.83 	& 1.60 	& 5.52 	& 1.83 \\
OrionA S 2 	& 05h39m37s 	& -07d24m32s 	& 5.58 	& 0.23 	& 4.26 	& 1.95 	& 6.46 	& 2.97 \\
OrionA S 3 	& 05h39m19s 	& -07d24m15s 	& 36.26 	& 0.33 	& 1.09 	& 0.84 	& 1.28 	& 0.98 \\
OrionA S 4 	& 05h40m02s 	& -07d28m15s 	& 19.81 	& 0.19 	& 2.03 	& 0.90 	& 2.33 	& 1.04 \\
OrionA S 5 	& 05h40m27s 	& -07d37m43s 	& 20.12 	& 0.23 	& 1.32 	& 0.76 	& 1.43 	& 0.82 \\
OrionA S 6 	& 05h40m30s 	& -07d44m18s 	& 14.89 	& 0.17 	& 1.02 	& 0.89 	& 1.26 	& 1.10 \\
OrionB NGC2023-2024 1 	& 05h41m27s 	& -01d43m38s 	& 0.72 	& 0.07 	& 14.87 	& 14.05 	& 15.51 	& 14.65 \\
OrionB NGC2023-2024 2 	& 05h41m17s 	& -01d47m55s 	& 0.76 	& 0.08 	& 11.70 	& 11.07 	& 12.74 	& 12.06 \\
OrionB NGC2023-2024 3 	& 05h41m49s 	& -01d55m38s 	& 30.73 	& 0.10 	& 1.22 	& 1.16 	& 1.65 	& 1.56 \\
OrionB NGC2023-2024 4 	& 05h41m51s 	& -01d58m00s 	& 73.42 	& 0.11 	& 0.79 	& 0.76 	& 1.45 	& 1.40 \\
OrionB NGC2023-2024 5 	& 05h41m56s 	& -02d01m10s 	& 1.22 	& 0.09 	& 8.51 	& 7.57 	& 9.49 	& 8.44 \\
OrionB NGC2023-2024 6 	& 05h41m38s 	& -02d19m08s 	& 50.29 	& 0.32 	& 1.97 	& 1.30 	& 2.00 	& 1.32 \\
OrionB NGC2023-2024 7 	& 05h41m37s 	& -02d25m49s 	& 4.42 	& 0.15 	& 7.77 	& 6.82 	& 7.89 	& 6.94 \\
OrionB NGC2068-2071 1	& 05h46m10s 	& -00d16m31s 	& 8.35 	& 0.18 	& 2.86 	& 2.67 	& 3.73 	& 3.49
\enddata
 \tablecomments{\newtext  RA and DEC values quoted refer to the geometric center of each clump. Asterisks denote clumps whose maps include regions missing NH$_3$ data, most of which have very small covering fraction.
 \label{table:ClumpAlphas} 
 }
\end{deluxetable*}

\begin{figure*}[ht]
\centering
$\begin{array}{c}
	%\vspace{0.5cm}
    \hspace{0.0cm}
    \includegraphics[scale=0.75]{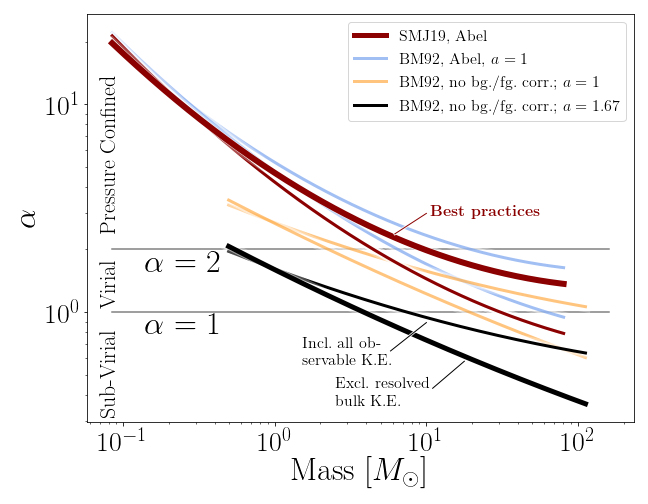}
\end{array}$
\caption{Estimates for $\alphacl$, compared.  We present only quadratic fits, suppressing individual clump data for clarity; each fit is plotted only over the appropriate range of $\Mcl$. In each line pair, $\calTbulk$ is included on the top line and excluded on the bottom line. Analysis choices can lead $\alphacl$ being underestimated by a factor of up to 3.5 at $\Mcl=100\,M_\odot$, relative to $\alphatwod$ with Abel reconstruction and including $\calTbulk$ (our recommended best-practices evaluation). }  
\label{fig:cascading_errors}
\end{figure*}

%%-------------- CONCLUSION --------------%%%
\section{Discussion} \label{S:conclusion}

{

As first stressed by \citet{Kauffmann2013}, many  studies have found massive clumps to be moderately to strongly sub-virial \citep{RomanDuval10_RingGMCprops, Pillai11_ICs, Wienen12_InnerGalaxyClumps, Ragan12_irdcs, Li13_MassiveOrionCores, Tan13_starless, Urquhart14_ATLASGALclumps, Friesen2016, Svoboda16_BolocamClumps, Kirk17_underPres, Keown2017,Redaellli17_B59, Csengeri17_atlasgal, Traficante18_Massive70micronClumpMotions,  Keown19Keystone, Chen19droplets, Kerr19virial, Billington20_atlasgalMaserClumps, Traficante20_Multiscale}. In our fiducial analysis, which follows SMJ19, we do not.  We trace the difference to several key features. 

First, we account for the component of kinetic energy associated with  spatial variations of the line-of-sight velocity in the plane of the sky, which we call  $\calTbulk$.  BM92 include this component (see their Appendix C), but many later studies  exclude it by adopting a typical value of $\sigma_{\rm eff}(x,y)$ for $\sigmacl$ in the evaluation of $\alphaBM$. There do exist exceptions, such as  \citet{RomanDuval10_RingGMCprops}, who include bulk energy in their equation (7), and works like   \citet{Colombo19_IntegratedGMCproperties} in which $\sigmaclz$ is defined in terms of the intensity-weighted variance of velocity (as adopted, for instance, by \citealt{Solomon87}).  Including $\calTbulk$ matters most for massive clumps: these contain supersonic internal motions, but their dense substructures tend to have narrow internal line widths with subsonic non-thermal components.  When the clump is well resolved, these substructures will tend to set the line width within each beam. The effect therefore varies with resolution.  This has been observed, for instance, in B5 by \citet{Pineda10_SharpTransition, Pineda15_QuadrupleStarSystem} and in Orion by \citet{Friesen2017} and \citet{Monsch18_OMC1_filament}. 

Second, we compute the gravitational self-energies of our clumps using the robust method of SMJ19.  Applied to an ensemble of clumps, this provides a sensitive calibration of the BM92 parameter $a$. An average over our clump sample indicates $a\simeq 1.13$, while fits to the mass dependence indicate that $a$ grows from unity for low clump masses to $\sim$1.15 for massive clumps.  {\newtext Because our clumps have generally modest projected aspect ratios (with the exception of HC2-2),} these represent low values of central concentration: the low-mass result $a\simeq1$ is consistent with these cores being essentially uniform-density, pressure-confined objects, and our entire sample is less centrally concentrated than a critical Bonnor-Ebert sphere (for which $a=1.22$).  Other analyses often ignore $a$, effectively taking $a=1$ (e.g., \citealt{Billington20_atlasgalMaserClumps});  or they adopt a specific value, such as  $a=1.25$ (corresponding to $k=3/2$: e.g., \citealt{Keown19Keystone}) or $a=1.67$ (corresponding to $k=2$: e.g., \citealt{RomanDuval10_RingGMCprops,Csengeri17_atlasgal}).  We note that  using the definition of $M_{\rm vir}$ in \citet{RohlfsWilson04} to estimate $\alphacl=M_{\rm vir}/\Mcl$  \citep[e.g.,][]{Wienen12_InnerGalaxyClumps} amounts to adopting $a=5/6$.  {\newtext We also note that our sample does not support the suggestion by \citet{BallesterosParedes18_OutOfVirial} that cases in which clumps are inferred to be unbound ($\alphacl>2$) might result from their values of $a$ being significantly underestimated.} 

Third, while we use dust emission to trace mass, we are careful to account for foreground and background emission because overestimating  $\Mcl$ leads one to underestimate $\alphacl$, roughly according to the scaling $\alphacl\propto \Mcl^{-1}$.  For this we prefer the Abel reconstruction method advanced in SMJ19.  

On this point, we note that prior studies determine clump masses and profiles using a wide range of data sources and analysis techniques \citep[see][]{Kauffmann2013}, not all of which exclude foreground and background contamination.  Common approaches include subtracting an interpolated background, matching and extracting a template profile, or making no correction. Dendrogram analysis \citep{Rosolowsky2008dendrograms} can employ any of these, although by default it makes no correction.  Approaches tailored to extract compact cores tend to fit templates \citep[e.g.,][]{Johnstone00_BEclumps} or subtract an estimated background, as in \citet{Menshchikov12_Getsources}.  As we mentioned in \S~\ref{S:method}, the spatial filtering afforded by interferometric or chopped observations accomplishes background subtraction in an approximate way. 

In Figure \ref{fig:cascading_errors} we review the effect of each of the three key features of our analysis mentioned above. We note that while  foreground/background subtraction can dominate the error for an individual clump, this effect is partly degenerate with the trend in $\alphacl(\Mcl)$, as shown in Figure \ref{fig:abel_edge}. Focusing on the mass scale $100\,M_\odot$, we see that neglecting $\calTbulk$ depresses one's estimate for $\alphatwod$ by about a factor of 1.8 (depending slightly on the clump extraction method). Including foreground and background emission suppresses it by a further factor of 1.4.   Adopting the extreme choice $a=1.67$ over the more accurate value of $1.15$ reduces $\alphacl$ further by a factor of $1.45$.  Compounding these leads to an underestimate of $\alphacl$ by a factor of 3.6.    

This analysis indicates that, while all three factors have comparable impact,  bulk kinetic energy is the most important 
effect at the reference mass of 100\,$M_\odot$, and the only 
one that is growing more important with increasing clump mass.  Although its impact is clearly resolution-dependent, this behaviour indicates that bulk kinetic energy might dominate at masses higher than those included in our sample. 

{\newtext A further and potentially significant difference is that we associate each clump with a definite  boundary on the sky, which we derive from the emission of a reliable dense gas tracer (NH$_3$) and which we use to infer quantities like the clump radius and gravitational energy. Compared to approaches in which radii are defined by moments of the emission, or by fitting to predetermined model templates, ours has the advantage that the derived quantities reflect those of the three-dimensional clump in question in a mathematically rigorous way (equations \ref{eq:avgEkin} and \ref{eq:avgEgrav}).   } 

In summary, we conclude that it is easy to underestimate $\alphacl$ by a significant margin as a result of choices in the analysis. Addressing these points carefully, as we do here, alleviates the physical puzzle that a finding of consistently sub-virial clumps would pose -- namely, that while a mildly sub-virial population may be close to equilibrium \citep{McKeeZweibel95, Tan13_starless} for the observed degree of magnetization \citep{Crutcher2012},  this is not true of the strongly sub-virial state. And, although one could prepare initial conditions far from  equilibrium in which  $\alphacl\ll1$, this state would be ephemeral, persisting only for a fraction of a free-fall time, as noted, for instance, by \citet{Kauffmann2013}. Moreover, it is not clear how a dense clump could be assembled in this state without concomitantly developing velocities at the virial scale. 

Within our Gould Belt sample, we find that clump motions are, in fact, virial in magnitude,  with $\alphacl \simeq 1.34$ and little dependence on mass at the upper end of our sample. This impacts any conclusions that depend on the virial ratio, such as the expected mode of protostellar accretion.  It remains to be seen whether conclusions drawn from this sample will be relevant to samples that include more massive clumps. 

{\newtext Our result that the virial parameter levels off at a value of order unity, rather than continuing to decline with increasing clump mass, points to a change in physical or evolutionary state for the most massive clumps in our sample. Could this reflect a greater role for dynamical feedback from star formation?  This would be consistent with the fact that protostellar outflows are active in a number of our massive clumps \citep{Kun08_CepheusSF,Devine09_L1228_HHobj,Curtis10_PerseusOutflows,Arce10_CompleteOFsInPerseus,Feddersen20_OrionOutflows}.  It is also consistent with the fact that class 0 protostars are highly segregated toward these massive clumps and the structures that host them, as is evident in Orion A \citep[e.g.,][]{Stutz15_OrionProtostarsAndNh} and Perseus \citep[e.g.,][]{Mercimek17_PerseusSFR}.  

It is already well established that the local star formation rate has a threshold-like \citep{Onishi98_Thresholds} or power law \citep{Heiderman10_KSrelations} dependence on $\Sigma$, and that the recent history of star formation correlates with the slope of the local probability density function of $\Sigma$ \citep{Sadavoy13_PhD,Stutz15_OrionProtostarsAndNh}. We find that the clump-averaged values of $\Sigma$ and $\Sigmacl$ both correlate with $\Mcl$ and anti-correlate with $\alphacl$, so the former rule, at least, is qualitatively consistent with and anti-correlation between $\alphacl$ and star formation intensity.  We consider this a promising avenue for future work.  

}

\acknowledgements We thank our referee and Christopher McKee for useful feedback and important suggestions.  A.S. thanks Tina Peters for advice on data analysis techniques. The research of A.S. and C.D.M. is supported by an NSERC Discovery Grant. AP acknowledges the support from the Russian Ministry of Science and Higher Education via the State Assignment Project FEUZ-2020-0038. AP is a member of the Max Planck Partner Group at the Ural Federal University.  We acknowledge that this work was conducted on the traditional land of the Huron-Wendat, the Seneca, and the Mississaugas of the Credit; we are grateful for the opportunity to work here. 

\software{Astropy \citep{astropy:2013, astropy:2018}, Scikit-learn \citep{scikit-learn} }

\newpage
%%%-------------- APPENDIX --------------%%%
\begin{appendix}
{\newtext
We describe here our procedure for managing discretization errors in the computation of $\calWtwod$. 

SMJ19 suggest computing $\calWtwod$ from $\Sigmacl(x,y)$ via 
\begin{equation} 
\calWtwod = \frac{1}{\pi} \int \Psi(x,y) \Sigmacl(x,y)\, dx\,dy 
\end{equation} 

where 

\begin{equation} 
\Psi(x,y) = -\frac{G}{\sqrt{x^2+y^2}} \circledast \Sigmacl(x,y). 
\end{equation} 

This procedure is perfectly valid in the continuous limit, but requires a small modification to account for the fact that $\Sigmacl$ is sampled on a discrete grid at some resolution $\delta x$.  The most important consequence of discreteness is that, because the Green's function $-G/\sqrt{x^2+y^2}$ diverges at the origin, its central element must be replaced with some finite value $-\varepsilon G/\delta x$, where $\varepsilon$ parameterizes the self-energy of a single pixel.   We adopt 

\begin{equation}
\varepsilon = 4\left[ \sinh^{-1}(1) - \frac{\sqrt{2}-1}3 \right] = 2.973,
\end{equation} 

the value that correctly yields the self-energy of a square of width $\delta x$ in the case that $\Sigmacl(x,y)$ is uniform.   We test this choice using discretized projections of structures of known self-gravitational energy, such as uniform-density spheres, and find that it reduces the error in $\calWtwod$ by a factor of about 2.4 relative to the outcome of choosing $\varepsilon=0$.   Significantly smaller improvements can be obtained by adjusting the non-central elements of the Green's function to account for the finite sizes of pixels.  However, we do not implement these because the error can further be  reduced by interpolating $\Sigmacl$ to a finer grid before evaluating $\calWtwod$.   Finding that the relative error when using $\varepsilon$ is $\sim \delta x/2\Rcl$, we adopt a strategy in which each clump's column density is re-sampled  so that $\Rcl> 100 \delta x$, implying errors well below one percent. 
} 
\end{appendix}
}
%%%-------------- REFERENCE --------------%%%
\bibliographystyle{apj} 
\bibliography{Virial_Parameter}

\end{document}